\documentclass[prd,showpacs,twocolumn,floatfix,nofootinbib]{revtex4-1}
\usepackage[dvipdfmx]{graphicx}
\usepackage{amsmath,amssymb,natbib,siunitx,color}

\begin{document}

\title{Gravitational-wave cosmography with LISA and the Hubble tension}

\author{Koutarou Kyutoku$^{1,2,3}$ and Naoki Seto$^4$}
\affiliation{
$^1$Theory Center, Institute of Particle and Nuclear Studies, KEK,
Tsukuba 305-0801, Japan\\
$^2$Department of Particle and Nuclear Physics, the Graduate University
for Advanced Studies (Sokendai), Tsukuba 305-0801, Japan\\
$^3$Interdisciplinary Theoretical Science (iTHES) Research Group, RIKEN,
Wako, Saitama 351-0198, Japan\\
$^4$Department of Physics, Kyoto University, Kyoto 606-8502, Japan
}

\date{\today}

\begin{abstract}
 We propose that stellar-mass binary black holes like GW150914 will
 become a tool to explore the local Universe within $\sim\SI{100}{Mpc}$
 in the era of the Laser Interferometer Space Antenna (LISA). High
 calibration accuracy and annual motion of LISA could enable us to
 localize up to $\approx60$ binaries more accurately than the error
 volume of $\approx\SI{100}{Mpc^3}$ without electromagnetic counterparts
 under moderately optimistic assumptions. This accuracy will give us a
 fair chance to determine the host object solely by gravitational
 waves. By combining the luminosity distance extracted from
 gravitational waves with the cosmological redshift determined from the
 host, the local value of the Hubble parameter will be determined up to
 a few \% without relying on the empirically constructed distance
 ladder. Gravitational-wave cosmography would pave the way for
 resolution of the disputed Hubble tension, where the local and global
 measurements disagree in the value of the Hubble parameter at
 $3.4\sigma$ level, which amounts to $\approx9\%$.
\end{abstract}

\maketitle

\section{Introduction} \label{sec:intro}

The discovery of GW150914, the first direct detection of gravitational
waves by Advanced LIGO, has significant impact on physics and astronomy
\cite{ligovirgo2016}. The key properties of the binary black holes
detected so far are the large masses and the high merger rate
\cite{ligovirgo2016-5}. In addition to the upcoming observations by
ground-based detectors, the prospects for observing binary black holes
at low frequency by space-borne detectors such as the Laser
Interferometer Space Antenna (LISA: see Ref.~\cite{lisapf2016} for LISA
pathfinder), called the evolved LISA (eLISA) for the last several years,
are attracting considerable attention
\cite{ligovirgo2016-2,sesana2016,seto2016,kyutoku_seto2016}.

Around $\SI{7}{\milli\hertz}$, LISA would be able to detect $\sim100$
nearly monochromatic binary black holes within $\sim\SI{100}{Mpc}$ at
high signal-to-noise ratios assuming the sensitivity of eLISA
\cite{kyutoku_seto2016}. This is remarkably different from binary
neutron stars, which had been widely considered to be the most promising
gravitational wave sources. Indeed, the expected detection number
($\propto\mathcal{M}^{10/3}$) and the maximum detectable distance
($\propto\mathcal{M}^{5/3}$) both depend strongly on the chirp mass
$\mathcal{M}$ of the nearly monochromatic binaries
\cite{kyutoku_seto2016}. Here, the chirp mass of the binary black hole
GW150914 is $28M_\odot$, whereas that of typical binary neutron stars is
only $\approx1.2M_\odot$. We should also note that the ground-based
detectors would detect no more than $O(1)$ events within
$\sim\SI{100}{Mpc}$ in a year unless the merger rate of compact binaries
is substantially higher than \SI{100}{Gpc^{-3}.yr^{-1}}.

Monochromatic binaries of massive black holes may serve as a useful tool
in astrometry (positional astronomy), which is the basis of astronomy
and astrophysics, because LISA will be able to localize them very
accurately \cite{kyutoku_seto2016}. Specifically, host objects (galaxies
or clusters) of many extragalactic gravitational-wave sources are likely
to be determined without observing electromagnetic counterparts that are
presumably absent for binary black holes \cite{lyutikov2016}. For this
purpose, LISA has two advantages compared to ground-based
detectors. First, high calibration accuracy of LISA will allow us to
determine the amplitude of gravitational waves and hence the luminosity
distance $D$ with negligible systematic errors \cite{cutler_holz2009},
in contrast to the ground-based detectors \cite{martynov_etal2016} (but
see Ref.~\cite{tuyenbev_etal2017} for improvement). Second, LISA's
annual motion induces a Doppler shift to the phase and modulation to the
amplitude, and thus the sky location can be determined accurately for
the long-lived sources \cite{cutler1998,takahashi_seto2002}. These two
advantages are combined to give a small error volume for the
stellar-mass binary black holes, which opens the possibility to
determine the host object of the binary.

In this paper, we discuss prospects for determining the Hubble parameter
in the local Universe by observing monochromatic stellar-mass binary
black holes with LISA. Although the Hubble parameter is arguably the
most fundamental quantity in cosmology, its value is still greatly
debated. In particular, it has recently been claimed that the Hubble
parameter determined by the local measurement,
$H_0=73.24\pm\SI{1.74}{km.s^{-1}.Mpc^{-1}}$ \cite{riess_etal2016}, is
larger by $3.4\sigma$ than that obtained from the cosmic microwave
background, $66.93\pm\SI{0.62}{km.s^{-1}.Mpc^{-1}}$
\cite{planck2016}. Local underdensity (cosmic void) seems to contribute
at some level to raise the local Hubble parameter \cite{jha_rk2007}, but
this effect does not account for the majority of $3.4\sigma$
\cite{marra_asv2013,bendayan_dms2014,wojtak_kwihryg2014}. While this
discrepancy, called the Hubble tension, could be reconciled by invoking
nonstandard cosmological ingredients such as dark radiation, it is
desirable to carefully analyze possible systematics associated with the
current measurements.

For this purpose, the gravitational-wave standard siren is a powerful
tool to study the local expansion rate of the Universe
\cite{schutz1986}. By observing binary black holes, we can determine the
luminosity distance from simple principles of physics and can also
calibrate the empirically constructed distance ladder at various
distance scales. However, it is well known that gravitational waves do
not tell us directly the cosmological redshift of binaries. Thus, the
host identification is crucial to extract redshift information.  Here,
binary black holes like GW150914 could now become promising standard
sirens to almost exclusively examine the distance scale around
\SI{100}{Mpc}, thanks to the exquisite localizability described above.

Following our previous work \cite{seto2016,kyutoku_seto2016}, the
fiducial values of chirp mass, $\mathcal{M}$, and comoving merger rate,
$R$, are taken to be optimistic values of $28M_\odot$ and
\SI{100}{Gpc^{-3}.yr^{-1}}, respectively (see
Ref.~\cite{ligovirgo2016-5} for an update). Recall this chirp mass is
larger by a factor of $>20$ than that for typical binary neutron
stars. We take the fiducial value of the Hubble parameter,
$H_0=h\times\SI{100}{km^.s^{-1}.Mpc^{-1}}$, to be $h=0.7$. Because we
focus on the local Universe in this study, we neglect all minor
corrections associated with the cosmological redshift.

\section{Localization of binary black holes with LISA} \label{sec:local}

We first describe the prospects for localizing stellar-mass binary black
holes by LISA based on Ref.~\cite{kyutoku_seto2016}. In this study, we
focus on the eLISA's N2A5 configuration \cite{klein_etal2016} as the
optimistic sensitivity of LISA and set the fiducial observation period
$T$ to be \SI{3}{yr}. We only consider the four-link (two-arm)
configuration and likely restoration of the six-link (three-arm)
configuration is highly welcome. The dependence of our results on the
noise curve and observation period is briefly described in
Sec.~\ref{sec:number}. Because we discuss the localization accuracy
using the result of Ref.~\cite{takahashi_seto2002} derived by a
quasimonochromatic approximation where nearly monochromatic but slightly
chirping binaries are considered, we restrict our attention to binaries
that do not merge within $T$. Namely, we focus on
$f<f_\mathrm{merge}(T)$, where
\begin{equation}
 f_\mathrm{merge} (T) = \SI{19.2}{\milli\hertz} \left(
                                                 \frac{\mathcal{M}}{28M_\odot}
                                                \right)^{-5/8} \left(
                                                \frac{T}{\SI{3}{yr}}
                                                               \right)^{-3/8}
                                                .
\end{equation}
This frequency range contains the majority of detectable binaries, and
high localization accuracy is also achieved in this range due to the
high signal-to-noise ratio \cite{kyutoku_seto2016}. In this
approximation, the signal-to-noise ratio for a binary at frequency $f$
is given by \cite{kyutoku_seto2016}
\begin{align}
 \rho(f) & = 21 \left( \frac{\mathcal{M}}{28M_\odot} \right)^{5/3}
 \left( \frac{T}{\SI{3}{yr}} \right)^{1/2} \left(
 \frac{D}{\SI{100}{Mpc}} \right)^{-1} \notag \\
 & \times \left( \frac{S(f)}{\SI{2.3e-41}{\per\hertz}} \right)^{-1/2}
 \left( \frac{f}{\SI{7}{\milli\hertz}} \right)^{2/3} , \label{eq:snr}
\end{align}
where $S(f)$ is the noise spectral density of LISA/eLISA
\cite{klein_etal2016}. We normalize the frequency and noise spectral
density (N2A5 configuration) by \SI{7}{\milli\hertz}, because this turns
out later to be the frequency that contributes most to the number of
accurately localized sources.

The expected errors of the angular position and luminosity distance are
given by \cite{takahashi_seto2002}
\begin{align}
 \Delta \Omega (f) & \sim \SI{3.6e-4}{str} \left( \frac{\rho}{20}
 \right)^{-2} \left( \frac{f}{\SI{7}{\milli\hertz}} \right)^{-2} , \\
 \frac{\Delta D}{D} & \sim 0.1 \left( \frac{\rho}{20} \right)^{-1} ,
 \label{eq:diserr}
\end{align}
and hence the error volume $\Delta V\equiv(4/3)D^2\Delta\Omega\Delta D$
is estimated to be
\begin{equation}
 \Delta V (f) \sim \SI{50}{Mpc^3} \left( \frac{D}{\SI{100}{Mpc}}
                                  \right)^3 \left( \frac{\rho}{20}
                                            \right)^{-3} \left(
                                            \frac{f}{\SI{7}{\milli\hertz}}
                                                         \right)^{-2}
                                            . \label{eq:evs}
\end{equation}
It has to be cautioned that this expression is derived in a conservative
manner by averaging over the inclination and sky location of the
binary. The precise size of the error volume depends on these
parameters, and we only discuss the averaged behavior. In addition, this
expression is derived by the Fisher analysis, where a high
signal-to-noise ratio is assumed. Although our primary targets are
binaries with strong signals that can be localized accurately, this
needs future refinement.

The peculiar velocity $\delta v$ of the host object has to be taken into
account for the identification in a redshift catalog of galaxies. From
the luminosity distance, $D$, obtained by LISA, we can approximately
estimate the redshift of the host by $H_0D/c$ with $c$ the speed of
light, assuming the Hubble expansion with a speculated value of
$H_0$. However, the actual redshift of the host has an additional drift
$\delta v/c$ induced by its peculiar velocity. Thus, for the host
identification in a redshift catalog, we need to increase the redshift
range by $\pm\sigma/c$, where $\sigma$ is a typical velocity dispersion
of galaxies. We can effectively handle this effect by using the total
distance error $\sqrt{(\Delta D)^2 +(\sigma/H_0)^2}$ for calculating
$\Delta V$ instead of the original one, $\Delta D$. In this study, we
conservatively take the fiducial value of $\sigma$ to be
$\SI{e3}{\kilo\meter\per\second}$ \cite{strauss_willick1995}, and
$\sigma/H_0=\SI{14}{Mpc}(\sigma/\SI{e3}{\kilo\meter\per\second})(h/0.7)^{-1}$. Alternatively,
we may regard this value as a sum of a more typical peculiar velocity
and the uncertainty in the value of $H_0$. This term dominates the
distance error for high signal-to-noise ratios, and the error volume is
given by
\begin{align}
 \Delta V(f) & \sim \SI{70}{Mpc^3} \left( \frac{D}{\SI{100}{Mpc}}
 \right)^2 \left( \frac{\rho}{20} \right)^{-2} \notag \\
 & \times \left( \frac{f}{\SI{7}{\milli\hertz}} \right)^{-2} \left(
 \frac{\sigma}{\SI{e3}{\kilo\meter\per\second}} \right) \left(
 \frac{h}{0.7} \right)^{-1} \label{eq:evp}
\end{align}
in the limit where $\Delta D/D$ is negligible.

For a given binary, the error volume depends on LISA's design as
$[S(f)/T]^{3/2}$ for large errors [Eq.~\eqref{eq:evs}] and as $S(f)/T$
for small errors [Eq.~\eqref{eq:evp}] via the signal-to-noise
ratio. This indicates that both the high sensitivity and the long-term
operation will be helpful for promoting monochromatic binaries to useful
standard sirens.

Once the best estimate of $H_0$ is obtained (e.g., by binary-black-hole
standard sirens), the peculiar velocity may be reversely determined for
individual galaxies. The peculiar velocity field could also serve as a
useful cosmological probe \cite{strauss_willick1995}.

\section{Number of accurately localized binaries} \label{sec:number}

A crucial question for gravitational-wave cosmography is how many
standard sirens are localized to sufficient accuracy. The number density
of binary black holes in each logarithmic frequency interval is given by
\cite{kyutoku_seto2016}
\begin{align}
 \frac{dn(f)}{d \ln f} & = \SI{1.2e-5}{Mpc^{-3}} \left(
 \frac{\mathcal{M}}{28M_\odot} \right)^{-5/3} \notag \\
 & \times \left( \frac{R}{\SI{100}{Gpc^{-3}.yr^{-1}}} \right) \left(
 \frac{f}{\SI{7}{\milli\hertz}} \right)^{-8/3} .
\end{align}
Combining this with Eqs.~\eqref{eq:evs} or \eqref{eq:evp}, the number of
binary black holes localized more accurately than a given error volume
of $\Delta V$ in each logarithmic frequency interval is deduced to be
\begin{widetext}
 \begin{align}
  \frac{dN(<\Delta V;f)}{d\ln f} & \sim
  \begin{cases}
   80 \; (\Delta V/\SI{100}{Mpc^3})^{1/2}
   (f/\SI{7}{\milli\hertz})^{-2/3} & (\Delta D \gg \sigma /H_0) \\
   70 \; (\Delta V/\SI{100}{Mpc^3})^{3/4}
   (f/\SI{7}{\milli\hertz})^{-1/6} (\sigma
   /\SI{e3}{\kilo\meter\per\second})^{-3/4} (h/0.7)^{3/4} & (\Delta D
   \ll \sigma /H_0)
  \end{cases}
  \notag \\
  & \times \left( \frac{\mathcal{M}}{28M_\odot} \right)^{5/6} \left(
  \frac{T}{\SI{3}{yr}} \right)^{3/4} \left(
  \frac{S(f)}{\SI{2.3e-41}{\per\hertz}} \right)^{-3/4} \left(
  \frac{R}{\SI{100}{Gpc^{-3}.yr^{-1}}} \right) .
 \end{align}
\end{widetext}
The smaller one of the above two gives a reasonable estimate, and the
transition occurs when $\Delta D=\sigma/H_0$ is satisfied.

Important information extracted from the above expressions may be
summarized as follows. By quantifying the frequency dependence (see
Ref.~\cite{kyutoku_seto2016} for a graphical representation),
$f\sim\SI{7}{\milli\hertz}$ is found to be abundant in accurately
localized binaries for typical configurations of eLISA, and $\Delta
V\lesssim\SI{100}{Mpc^3}$ is achieved at \SI{100}{Mpc} around
\SI{7}{\milli\hertz} for both type of errors. The number of accurately
localized binaries is proportional to $\mathcal{M}^{5/6}$ for both
cases. Stated differently, the number will be determined by the weighted
average, $\langle\mathcal{M}^{5/6}\rangle$. As for LISA's design, the
above expression indicates that the number is proportional to
$[T/S(f)]^{3/4}$. In fact, the realistic increase in time will be faster
than $T^{3/4}$, because longer observations will improve the detector
sensitivity by removing more foreground noise associated with Galactic
binary white dwarfs, so that $S(f)$ will be reduced simultaneously
\cite{klein_etal2016}.

\begin{figure}
 \includegraphics[width=.95\linewidth]{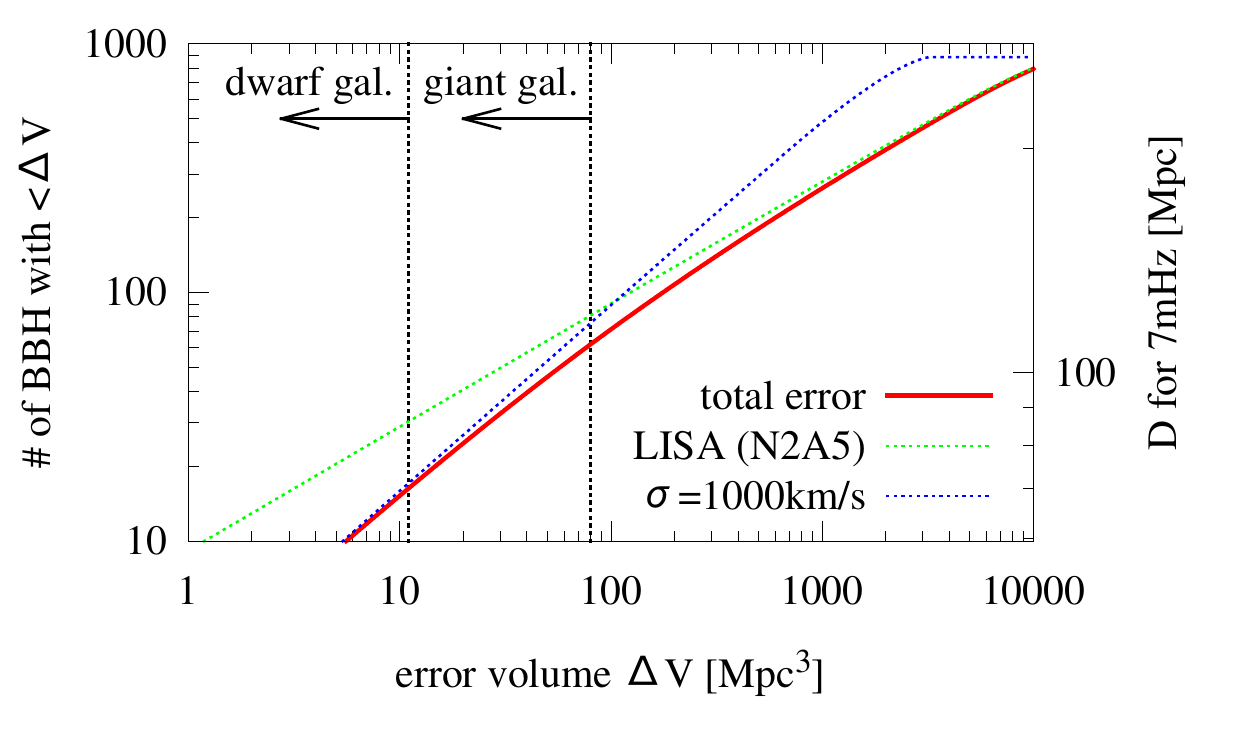}
 \caption{Cumulative number of nonmerging binary black holes with the
 error volume smaller than $\Delta V$ (the red solid curve). The green
 and blue dashed curves are asymptotic relations derived using the
 distance error associate with only the statistical error of LISA and
 only the peculiar velocity of galaxies, respectively. The blue curve is
 saturated at $\sim900$, which corresponds to the number of binaries
 with $\rho>8$ (hypothetical detection threshold). The vertical dashed
 lines denote the inverse of typical number densities of dwarf (left)
 and giant (right) galaxies \cite{baldry_etal2012}. The right axis shows
 the luminosity distance corresponding to the number of binaries (left
 axis) under the assumption that all the nonmerging binaries are
 uniformly filling the effective volume for \SI{7}{\milli\hertz}.}
 \label{fig:number}
\end{figure}

Figure \ref{fig:number} shows the total number of nonmerging binary
black holes localized more accurately than $\Delta V$, $N(<\Delta
V)$. The estimate derived by the total distance error is presented as a
red curve, and we also show asymptotic relations for cases that one of
the distance error becomes dominant. This figure shows that $N(<\Delta
V)$ integrated over frequency follows the expected relation,
$\propto(\Delta V)^{3/4}$ at small $\Delta V$ and $\propto(\Delta
V)^{1/2}$ at large $\Delta V$, until it levels off at the detection
threshold. For comparison, we indicate the volume corresponding to the
typical number density of giant galaxies with $\gtrsim10^9 M_\odot$ and
dwarf galaxies with $\gtrsim10^7M_\odot$. In this study, we adopt a
galaxy stellar mass function of Ref.~\cite{baldry_etal2012} derived by
the GAMA survey in the local Universe for concreteness, and any
reasonable function does not modify our discussions.

\section{Host galaxy determination} \label{sec:host}

The measurement error of the Hubble parameter, $H_0=cz/D$, is determined
by the statistical error of the luminosity distance and contamination of
the cosmological redshift due to the peculiar velocity. The cosmic
variance due to the coherent peculiar velocity field is non-negligible
in the local measurement concerned here, and its estimation is an
entirely different topic from our current problem
\cite{marra_asv2013,bendayan_dms2014,wojtak_kwihryg2014}. To the first
order in the measurement error, the sum of the distance error and the
contamination due to the random peculiar velocity can be approximated by
$\sqrt{(\Delta D/D)^2 +[\sigma/(H_0 D)]^2}$. This term behaves as
$\propto D^{-1}$ at the small distance and $\propto D$ at the large
distance. As the number $N\propto D^3$ of available standard sirens
increases, the error reduces approximately as $N^{-1/2}\propto
D^{-3/2}$. Thus, the measurement error of the Hubble parameter is
conservatively estimated by $(\Delta D/D)/\sqrt{N}$ using
Eq.~\eqref{eq:diserr} for the farthest standard siren. The problem is
how many binary black holes can be utilized as the standard sirens, and
the number depends crucially on how many of them can be associated with
the unique host.

We first consider the scenario that the majority of binary black holes
are associated with giant galaxies with $\gtrsim10^9M_\odot$. This seems
plausible, because more than 90\% of the galaxy stellar mass is
contained in giant galaxies \cite{baldry_etal2012}. In this scenario,
$\approx60$ binaries at $\lesssim\SI{100}{Mpc}$ may be assigned with
unique host galaxies. This value is quite promising for determining the
local Hubble parameter, because the error in the luminosity distance and
the scatter associated with the random peculiar velocity are suppressed
as $1/\sqrt{N(<\Delta V)}$ and may become subdominant
\cite{shi_turner1998}. Then, the cosmic variance will limit the accuracy
to a few \% \cite{shi_turner1998}, which is better than the extent of
the current Hubble tension \cite{riess_etal2016}. The cosmic variance
becomes also irrelevant when our aim is to compare the local Hubble
parameters obtained by the traditional distance ladder and the
gravitational-wave cosmography, sampling an identical volume.

We should also note that most of the globular clusters are associated
with galaxies whose number density is $\approx\SI{0.01}{Mpc^{-3}}$
\cite{portegieszwart_mcmillan2000}, close to the giant galaxies
mentioned above. Thus, we can apply our discussions for giant galaxies
also to the dynamical formation scenario of binary black holes in
globular clusters.

The situation becomes subtle if binary black holes are primarily
associated with numerous dwarf galaxies with $\gtrsim10^7M_\odot$
\cite{lamberts_gch2016}. Taking the fact that the completeness of the
survey is not very high for dwarfs even in the local Universe, the
typical number density should be no lower than
$\sim\SI{0.1}{Mpc^{-3}}$. In this case, the number of binary black holes
with unique host galaxies is reduced to $\approx15$ located at
$\lesssim\SI{60}{Mpc}$ at best. Thus, the determination of the Hubble
parameter becomes inaccurate due to the error sources raised above.

Even in this case, LISA will offer fruitful information on the local
Universe. Gravitational-wave observations will help to calibrate the
distance ladder by providing accurate estimates of the luminosity
distance. Furthermore, the number will be sufficient to infer typical
host galaxies of stellar-mass binary black holes. Because the host
determination of binary black holes is extremely difficult with
ground-based detectors \cite{fairhurst2009,grover_ffmrsv2014},
investigation of host galaxies can be one of important scientific goals
of LISA.

Galaxies in the Local Group may also help us to understand the
characteristics of typical host galaxies of binary black holes. As
pointed out in Ref.~\cite{seto2016}, Milky Way equivalent galaxies
should typically possess massive binary black holes in the very low
frequency range of $\lesssim\SI{0.5}{\milli\hertz}$. Even though LISA's
sensitivity is low at very low frequency, the signal could be detected
with a moderate strength due to the proximity of the Local Group
including small galaxies such as the Small Magellanic Cloud. The
localization may be easy for such close galaxies.

Although we have restricted our attention to the binaries with unique
host candidates for simplicity, binaries with multiple host candidates
can also be utilized to extract cosmological information in a
statistical manner, e.g., by assigning appropriate likelihood for each
candidate in the error volume
\cite{macleod_hogan2008,petiteau_bs2011}. This strategy is appealing,
because the accessible distance and hence the number of available
binaries increase. A large number of dwarf galaxies can be handled in
the same manner. A similar strategy also works for the case that the
host galaxy exists in a relatively dense structure such as a cluster and
thus the error volume of $\approx\SI{100}{Mpc^3}$ or even
$\approx\SI{10}{Mpc^3}$ (see vertical lines in Fig.~\ref{fig:number}) is
not small enough to uniquely determine the host galaxy. In this case, we
can alternatively take the average redshift of the relevant galaxies,
resulting in reduction of the scatter induced by random motions within
the structure.

\section{Comparison with other proposals of gravitational-wave
 cosmology} \label{sec:other}

As stated in Sec.~\ref{sec:intro}, stellar-mass binary black holes will
serve as a unique standard siren to investigate the local Universe
around \SI{100}{Mpc}. Supermassive binary black holes have frequently
been discussed as a standard siren for eLISA/LISA (see, e.g.,
Ref.~\cite{holz_hughes2005,deffayet_menou2007,tamanini_cbskp2016,caprini_tamanini2016},
and see also Ref.~\cite{sereno_sbjvb2010,sereno_jsv2011} for an
alternative approach). Because targeted redshifts are usually
$z\gtrsim1$ for such sources, they will play an important role to bridge
the gap between local and global (i.e., cosmic microwave background)
measurements and also to explore the dark-energy equation of state. This
strategy, however, heavily relies on accurate localization by
electromagnetic counterparts, whose association is not guaranteed (but
see also Ref.~\cite{petiteau_bs2011} for statistical redshift
determination). Whereas extreme mass ratio inspirals may serve as a
probe to the relatively local Hubble parameter of $z\gtrsim0.1$, the
redshift has to be determined by statistical methods even with LISA
\cite{macleod_hogan2008}. In the future, the Big Bang Observer (BBO) may
become another important tool to study cosmology with ultrahigh
precision by observing compact binary coalescences from the space
\cite{cutler_holz2009,arabsalmani_ss2013}.

\section{Summary and future outlook} \label{sec:summary}

We proposed a possibility of novel gravitational-wave cosmography by
stellar-mass binary black holes with a space-borne detector, LISA. High
accuracy of the luminosity-distance and angular-position estimation by
LISA may allow us to determine host objects. We estimate that the local
Hubble parameter will be determined with accuracy of up to a few \% via
extraction of the cosmological redshift from the hosts. Because the
luminosity distance is determined absolutely, gravitational-wave
cosmography would give a clue to resolve the disputed Hubble tension
\cite{riess_etal2016}. At the same time, LISA will also enable us to
investigate host galaxies of binary black holes such as their types
\cite{lamberts_gch2016}.

The analysis performed in this study can be improved in various
directions. Examples include Bayesian parameter estimation allowing
multiple host galaxies, mock simulations using galaxy catalogs, and
careful assessment of the finite-number effect and cosmic variance. It
would also be interesting to examine the prospect for studying the
properties of individual galaxies and clusters of galaxies with binary
black holes. We leave these topics for future study.

\begin{acknowledgments}
 We would like to thank Juna A. Kollmeier and E. Sterl Phinney for
 useful discussions at Kavli Institute for Theoretical Physics (KITP)
 and Masato Onodera for fruitful discussions. This work is supported by
 the National Science Foundation (NSF) under Grant No. NSF PHY11-25915
 and by Japanese Society for the Promotion of Science (JSPS) Kakenhi
 Grant-in-Aid for Research Activity Start-up (No.~15H06857), for
 Scientific Research (No.~15K05075, No.~16H06342) and for Scientific
 Research on Innovative Areas (No.~24103006). Koutarou Kyutoku is
 supported by the RIKEN iTHES project.

 \textit{Note added.}---At the 11th LISA symposium we were informed that
 a related work (W. Del Pozzo and A. Sesana) was independently underway.
\end{acknowledgments}

\end{document}